\documentclass[twocolumn,showpacs,aps,prl,superscriptaddress,letterpaper]{revtex4}
\usepackage{graphicx}
\usepackage{dcolumn}
\usepackage{amsmath}
\usepackage{epsfig}

\RequirePackage{xspace}

\usepackage{relsize}
\def\babar{\mbox{\slshape B\kern-0.1em{\smaller A}\kern-0.1em
    B\kern-0.1em{\smaller A\kern-0.2em R}}}

\def\epem       {\ensuremath{e^+e^-}\xspace}

\def\taum       {\ensuremath{\tau^-}\xspace}

\def\nut        {\ensuremath{\nu_\tau}\xspace}

\def\Kbar  {\kern 0.2em\overline{\kern -0.2em K}{}\xspace}

\def\Kz    {\ensuremath{K^0}\xspace}
\def\Kzb   {\ensuremath{\Kbar^0}\xspace}
\def\KzKzb {\ensuremath{\Kz \kern -0.16em \Kzb}\xspace}
\def\Kp    {\ensuremath{K^+}\xspace}
\def\Km    {\ensuremath{K^-}\xspace}

\def\KpKm  {\ensuremath{\Kp \kern -0.16em \Km}\xspace}

\def\Dbar    {\kern 0.2em\overline{\kern -0.2em D}{}\xspace}

\def\Dz      {\ensuremath{D^0}\xspace}
\def\Dzb     {\ensuremath{\Dbar^0}\xspace}
\def\DzDzb   {\ensuremath{\Dz {\kern -0.16em \Dzb}}\xspace}
\def\Dp      {\ensuremath{D^+}\xspace}
\def\Dm      {\ensuremath{D^-}\xspace}

\def\DpDm    {\ensuremath{\Dp {\kern -0.16em \Dm}}\xspace}

\def\Bbar    {\kern 0.18em\overline{\kern -0.18em B}{}\xspace}

\def\Bz      {\ensuremath{B^0}\xspace}
\def\Bzb     {\ensuremath{\Bbar^0}\xspace}
\def\BzBzb   {\ensuremath{\Bz {\kern -0.16em \Bzb}}\xspace}
\def\Bu      {\ensuremath{B^+}\xspace}
\def\Bub     {\ensuremath{B^-}\xspace}

\def\BpBm    {\ensuremath{\Bu {\kern -0.16em \Bub}}\xspace}

\def\BorBbar    {\kern 0.18em\optbar{\kern -0.18em B}{}\xspace}
\def\DorDbar    {\kern 0.18em\optbar{\kern -0.18em D}{}\xspace}
\def\KorKbar    {\kern 0.18em\optbar{\kern -0.18em K}{}\xspace}

\mathchardef\Upsilon="7107
\def\Y#1S{\ensuremath{\Upsilon{(#1S)}}\xspace}

\def\FourS {\Y4S}

\mathchardef\Deltares="7101
\mathchardef\Xi="7104
\mathchardef\Lambda="7103
\mathchardef\Sigma="7106
\mathchardef\Omega="710A

\def\Deltabar{\kern 0.25em\overline{\kern -0.25em \Deltares}{}\xspace}
\def\Lbar{\kern 0.2em\overline{\kern -0.2em\Lambda\kern 0.05em}\kern-0.05em{}\xspace}
\def\Sigbar{\kern 0.2em\overline{\kern -0.2em \Sigma}{}\xspace}
\def\Xibar{\kern 0.2em\overline{\kern -0.2em \Xi}{}\xspace}
\def\Obar{\kern 0.2em\overline{\kern -0.2em \Omega}{}\xspace}
\def\Nbar{\kern 0.2em\overline{\kern -0.2em N}{}\xspace}
\def\Xb{\kern 0.2em\overline{\kern -0.2em X}{}\xspace}

\def\BR         {{\ensuremath{\cal B}\xspace}}

\def\pt         {\mbox{$p_T$}\xspace}

\newcommand{\tev}{\ensuremath{\mathrm{\,Te\kern -0.1em V}}\xspace}
\newcommand{\gev}{\ensuremath{\mathrm{\,Ge\kern -0.1em V}}\xspace}
\newcommand{\mev}{\ensuremath{\mathrm{\,Me\kern -0.1em V}}\xspace}
\newcommand{\kev}{\ensuremath{\mathrm{\,ke\kern -0.1em V}}\xspace}
\newcommand{\ev}{\ensuremath{\mathrm{\,e\kern -0.1em V}}\xspace}
\newcommand{\gevc}{\ensuremath{{\mathrm{\,Ge\kern -0.1em V\!/}c}}\xspace}
\newcommand{\mevc}{\ensuremath{{\mathrm{\,Me\kern -0.1em V\!/}c}}\xspace}
\newcommand{\gevcc}{\ensuremath{{\mathrm{\,Ge\kern -0.1em V\!/}c^2}}\xspace}
\newcommand{\mevcc}{\ensuremath{{\mathrm{\,Me\kern -0.1em V\!/}c^2}}\xspace}

\def\nb         {\ensuremath{{\rm \,nb}}\xspace}

\def\invfb   {\ensuremath{\mbox{\,fb}^{-1}}\xspace}

\def\mus  {\ensuremath{\rm \,\mus}\xspace}

\def\mus        {\ensuremath{\,\mu{\rm s}}\xspace}

\def\pep2{PEP-II}

\def\gsim{{~\raise.15em\hbox{$>$}\kern-.85em
          \lower.35em\hbox{$\sim$}~}\xspace}
\def\lsim{{~\raise.15em\hbox{$<$}\kern-.85em
          \lower.35em\hbox{$\sim$}~}\xspace}

\xspace

\def\geant      {\mbox{\tt GEANT}\xspace}

\def\jetset74   {\mbox{\tt Jetset \hspace{-0.5em}7.\hspace{-0.2em}4}\xspace}

\def\tausevenall {\taum \rightarrow  4\pi^- 3\pi^+(\pi^0)  \nut }
\def\tauseven {\taum \rightarrow  4\pi^- 3\pi^+ \nut }
\def\tausevenpinull {\taum \rightarrow  4\pi^- 3\pi^+\pi^0  \nut }

\def\eetoqq {e^+e^-\rightarrow q\bar q}

\newcommand{\BABARPubYear}    {05}
\newcommand{\BABARPubNumber}  {015}

\newcommand{\SLACPubNumber} {11229}

\newcommand{\lumi}    {232.2\invfb}

\def\figurebox#1#2#3{
    \def\arg{#3}
    \ifx\arg\empty
    {\hfill\vbox{\hsize#2\hrule\hbox to #2{\vrule\hfill\vbox to #1{\hsize#2\vfill}\vrule}\hrule}\hfill}%
    \else
    {\hfill\epsfbox{#3}\hfill}
    \fi}

\long\def\inst#1{\par\nobreak\kern 4pt\nobreak
    {\it #1}\par\vskip 10pt plus 3pt minus 3pt}

\begin{document}

\bibliographystyle{prsty}

\preprint{\babar-PUB-\BABARPubYear/\BABARPubNumber} 
\preprint{SLAC-PUB-\SLACPubNumber} 

%\begin{flushleft}
%\babar-PUB-\BABARPubYear/\BABARPubNumber\\
%SLAC-PUB-\SLACPubNumber\\
%hep-ex/\LANLNumber\\[10mm]
%\end{flushleft}

\title{
{\large \bf
Search for the Decay $\tausevenall$} 
}

\author{B.~Aubert}
\author{R.~Barate}
\author{D.~Boutigny}
\author{F.~Couderc}
\author{Y.~Karyotakis}
\author{J.~P.~Lees}
\author{V.~Poireau}
\author{V.~Tisserand}
\author{A.~Zghiche}
\affiliation{Laboratoire de Physique des Particules, F-74941 Annecy-le-Vieux, France }
\author{E.~Grauges}
\affiliation{IFAE, Universitat Autonoma de Barcelona, E-08193 Bellaterra, Barcelona, Spain }
% ...
\collaboration{The \babar\ Collaboration}
\noaffiliation

\date{\today}

\begin{abstract}
A search for the decay of the $\tau$ lepton to seven charged pions
and one or zero $\pi^{0}$ mesons was performed using the \babar\ detector at
the PEP-II asymmetric-energy $e^{+}e^{-}$ collider.
The analysis uses \lumi of data at center-of-mass energies on or near the
\FourS resonance.
We observe 24 events with an expected background of $21.6\pm1.3$ events.
Without evidence for a signal, we calculate an upper limit of
\BR$(\tausevenall) < 3.0\times10^{-7}$ at 90\,\% confidence level. 
This is an improvement by nearly an order of magnitude over the previously established limit.
In addition, we set upper limits for the exclusive decays $\tauseven$ and $\tausevenpinull$. 
\end{abstract}


\maketitle

The decay of the $\tau$ lepton to seven charged 
particles~\footnote{Throughout this paper, whenever a mode is given its charge conjugate is also implied.}
has not been observed to date.
An upper limit of 
$\BR(\tau^{-} \rightarrow 4 \pi^{-} 3 \pi^{+} (\pi^{0}) \nu_{\tau}) <
2.4\times10^{-6}$
at 90\,$\%$ confidence level has been set by the CLEO Collaboration~\cite{GAN}. 
Theoretical calculations using an effective chiral Lagrangian to estimate the matrix elements
show that the $\tau$ decay rate to seven charged particles is much smaller than the 
decay into five charged particles due to its smaller phase space.
This leads to a theoretical branching
fraction estimate of the order of $10^{-11}$, which could be enhanced
by up to an order of
magnitude if this decay proceeds via resonances~\cite{Theory}.

This analysis is based on data recorded 
by the \babar\ detector at the \pep2\ asymmetric-energy \epem storage 
ring operated at the Stanford Linear Accelerator Center.
The data sample consists of \lumi recorded at center-of-mass (CM) energies
of $10.58 \gev$ and $10.54 \gev$.
With an expected cross section for $\tau$ pairs for these CM energies 
of $\sigma_{\tau\tau} = (0.89\pm0.02)$ \nb \cite{kk},
the number of produced $\tau$ pair events $N_{\tau\tau}$ is $(206.6\pm5.2)\times 10^6$.\\

The \babar\ detector is described in detail in Ref.~\cite{detector} and
only a brief description is given here.
Charged-particle momenta are measured with a 5-layer
double-sided silicon vertex tracker (SVT) and a 40-layer drift chamber (DCH) 
inside a 1.5-T solenoidal magnetic field.
A calorimeter (EMC) consisting of 6580 CsI(Tl) 
crystals is used to measure electromagnetic energy.
A ring-imaging Cherenkov detector is used to identify
charged hadrons, in combination with ionization energy loss
measurements in the SVT and the DCH.
Muons are identified by an instrumented magnetic-flux return (IFR).
Particle four-momenta are reconstructed in the laboratory
frame and then boosted to the \epem\ center-of-mass (CM) 
frame using the measured beam energies.

We use Monte Carlo simulation techniques to estimate our signal efficiencies
and background contamination from other $\tau$-pair events.
The production of $\tau$ pairs is simulated with the {\mbox{\tt KK}\xspace} generator~\cite{KK2f},
and non-signal  $\tau$ lepton decays are modeled with {\mbox{\tt TAUOLA}\xspace}~\cite{tauola} according
to measured rates~\cite{pdg2002}.
Signal events were generated with uniform density throughout the
available phase space.
The simulation of the \babar\ detector is based on \geant~4~\cite{geant4}.\\

Events with eight charged tracks and a net charge of zero are selected.
Tracks are required to have 
a distance of closest approach to the interaction point in the
plane transverse to the beam axis (DOCA$\rm_{XY}$) of less than
1.5\,cm, and a distance of closest approach along the beam direction 
of less than 10\,cm.
It is required that at least five of these eight tracks have a
minimum transverse momentum (\pt) of $100\,\mevc$,
and 12 or more DCH hits.
Photons are reconstructed from EMC clusters and are required to have a minimum energy 
of $70\,\mev$, more than three crystal hits, 
and a lateral energy profile consistent with that of a photon.
The event is divided into hemispheres by a plane perpendicular to the thrust
axis~\cite{Sjostrand:1985ys}, where the thrust is calculated from all charged tracks and all photons in
the event. The event thrust has to be larger than 0.9.
We require all events to have one track in one hemisphere (the ``tag-side'' hemisphere)
and seven tracks in the other hemisphere (the ``signal side''
hemisphere). The above requirements define the 1-7 topology.\\

Monte Carlo studies have shown that the background from $\tau$-pair events stems from
photon conversions in the detector material.
To reduce this contribution and to exclude $\eetoqq$ ($q = \{u,d,s,c,b\}$)
decays containing kaons we apply particle identification on the signal side.
We demand that at least six of the tracks are identified as pions with
high probability.
We apply looser identification criteria to the seventh track.

To further reduce  photon conversions,
we require all seven tracks on the signal side to have a minimum transverse momentum of $100\,\mevc$
and the ratio of ${\rm DOCA_{XY}}/\pt$ to be less than $0.7\,{\rm cm}/(\gevc)$.

Further suppression of $\eetoqq$ events is achieved by requiring the tag side to satisfy 
one of the following criteria:
1) a tightly identified electron or muon with no more than one additional
photon on the tag-side;
2) a tight lepton veto and no additional photon; or
3) a reconstructed $\rho$ meson with an invariant mass of
$0.650< {\rm m}_{\rho} < 0.875\,\gevcc$,
derived from the combination of the
1-prong track with a reconstructed $\pi^0$ candidate.
The $\pi^0$ candidates are formed by combining two photons
and requiring an invariant mass between $0.113$ and $0.155\,\gevcc$.\\

\begin{figure}[tb]
\begin{center}
\epsfig{file=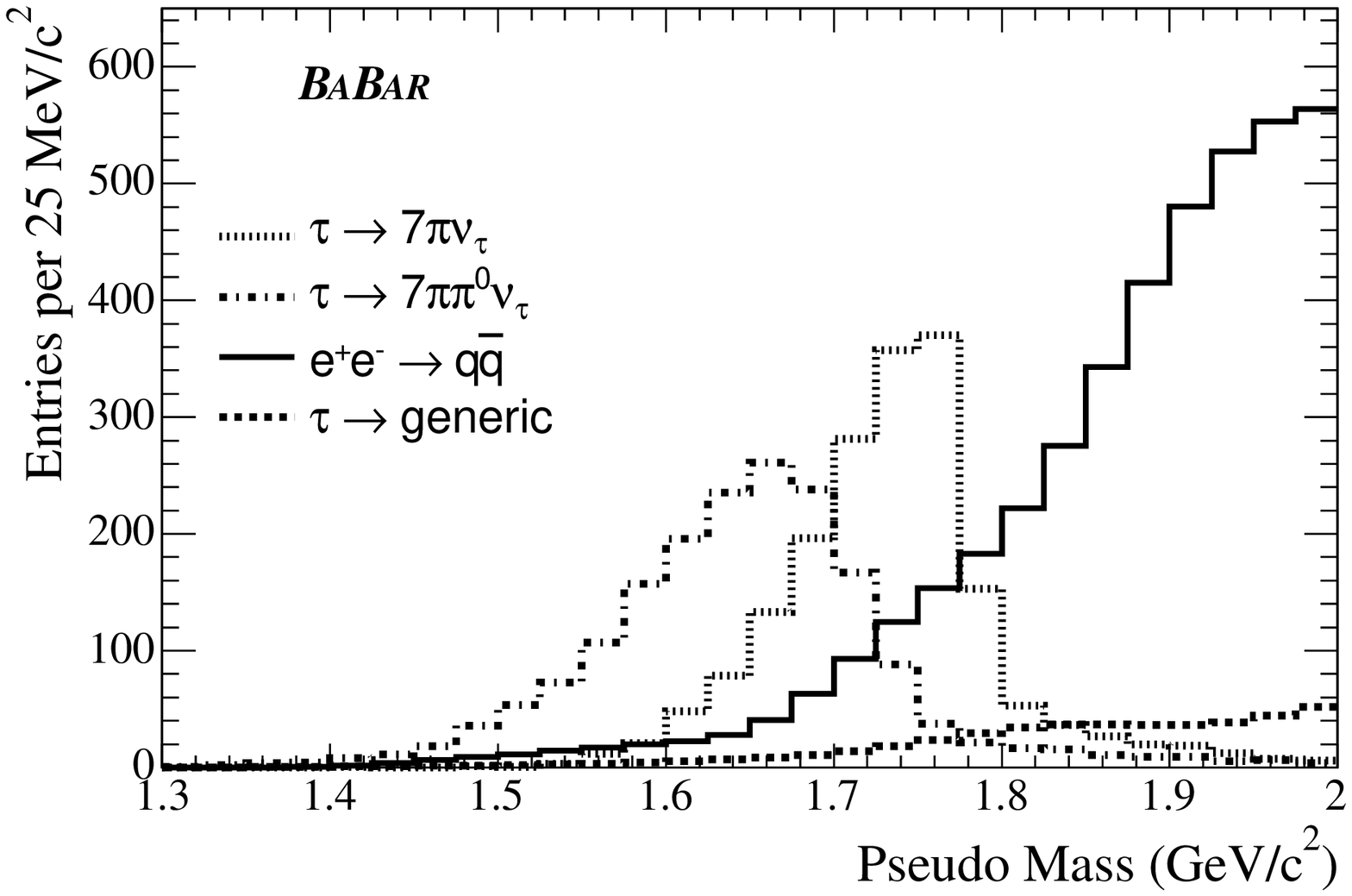,width=0.50\textwidth}
\end{center}
\caption{Pseudo mass distributions of the seven charged tracks after
  the 1-7 topology selection for signal (dashed-dotted:
$\tausevenpinull$; dotted: $\tauseven$), generic
  $\tau$ (including 5-prong decays) (dashed), and $\eetoqq$ Monte Carlo
events (solid). Entries are normalized to the data luminosity. To display
the two signal modes we assumed a branching fraction of $2\times10^{-5}$.\label{fig:mcmass}}
\end{figure}

After applying these requirements we calculate the pseudo mass $(m)$ of the $\tau$
lepton~\cite{pseudomass}: 
\begin{equation}
m^{2} = 2\,(E_{beam} - E_{7\pi})(E_{7\pi} - P_{7\pi}) +
m_{7\pi}^{2}. 
\end{equation}
The pseudo mass is an approximation 
of the invariant mass of the tau, where the neutrino's flight 
direction is approximated by the combined momentum vector $P_{7\pi}$ of the 
seven charged tracks, and its energy is taken to be the difference 
between the beam energy $E_{beam}$ in the center-of-mass system 
and the combined energy $E_{7\pi}$ of the charged tracks.
The pseudo mass allows for a better discrimination between signal
events and background from $\eetoqq$ events than $m_{7\pi}$.
The final event count is performed in the signal region
$1.3<m<1.8\,\gevcc$.
Figure~\ref{fig:mcmass} illustrates the pseudo mass spectra of
simulated signal and background contributions after the topology selection.\\

For this analysis, a comparison of Monte Carlo simulation and data has shown that
$\eetoqq$ background contributions cannot reliably be extracted from
simulation due to difficulties in modeling the fragmentation processes. 
The shape of the pseudo mass distribution appears to be correctly modeled, but the
overall normalization is not.
Therefore, we estimate this background directly from data, using a method 
described below.
On the other hand, the simulation of $\tau$ pair events yields a reliable
estimate of their expected background contribution, verified by 
loosening requirements that suppress the tau background.
A breakdown of the signal efficiencies and individual
$\tau$ background 
contributions are listed in Table~\ref{tab:eff} for each
selection step.
In this table the generic $\tau$ sample does not contain 5$\pi$ and 5$\pi\pi^{0}$
decays, which are listed separately since they comprise the only
background from $\tau$ decays after the final selection.

\begin{table}[tb]
\begin{center}
\caption{Cumulative signal efficiencies in \% and $\tau$ background
 contributions as number of events scaled
 to the data luminosity after the various selection criteria.
ID denotes Identification, Conv.~Conversion, and Gen.~Generic.\label{tab:eff}\\}
\begin{tabular}{l|c|c|c|c|c}
\hline
\hline
& \multicolumn{2}{c|}{Efficiency [\%]} & \multicolumn{3}{c}{Number of events} \\
 &{7$\pi\nu_{\tau}$}&{7$\pi\pi^{0}\nu_{\tau}$} & 
{{\small Gen.\,}$\tau$}&{5$\pi\nu_{\tau}$}&{5$\pi\pi^{0}\nu_{\tau}$} \\
\hline
1-7 Topology	& 23.6$\pm$1.4 & 22.1$\pm$1.4 & 767 & 198  & 187   \\
Particle ID	& 20.7$\pm$1.3 & 19.6$\pm$1.3 & 108 & 64   & 75    \\
Conv.\,Veto     & 15.8$\pm$1.0 & 14.9$\pm$1.0 &  0  & 4.7  & 9.2  \\
1-Prong\,Tag 	& 10.2$\pm$0.7  & 9.6$\pm$0.7 &  0  & 1.7  & 4.2   \\
Signal Region 	& 9.4$\pm$0.6   & 9.3$\pm$0.6 &  0  & 0.4  & 0.8   \\
\hline
\hline
\end{tabular}
\end{center}
\end{table}

The efficiencies for events in the signal region for 
$\tauseven$ and $\tausevenpinull$ are
$(9.4\pm0.1\pm0.6)\,\%$ and $(9.3\pm0.1\pm0.6)\,\%$ (see Table~\ref{tab:eff}).
The first errors are statistical and the second systematic.

The systematic errors on the signal efficiencies include contributions
from uncertainties in the
reconstruction of charged tracks (5.2\,\%), 
the uncertainty associated with the
particle identification on the signal and tag side (2.7\,\%), 
the luminosity measurement and the
$\tau$ pair cross-section determination (2.3\,\%), 
and the uncertainty on
the generic 1-prong $\tau$ decay branching ratios (0.5\,\%).
Since the efficiencies of the two decay
channels to the 7-prong $\tau$ decays are in good agreement, we average
them and obtain an overall efficiency for the decay $\tausevenall$ of
$(9.4\pm0.6)\,\%$ which includes the statistical and systematic contributions
described above.

To determine the number of $\eetoqq$ events in the signal region, 
we use the following procedure: we histogram the pseudo mass distribution 
of all data events that satisfy the topology requirement and are in the pseudo mass 
sideband $1.8 < m < 2.6\,\gevcc$. The contribution of the $\tau$ background, 
determined from the simulation, is subtracted from this histogram, and the 
resulting distribution is fit with a Gaussian function. We then fix the mean 
and the width of this Gaussian, and use it to fit the $\tau$-background-subtracted 
distributions of sideband events for all subsequent cuts, floating only the normalizations. 
Integrating the area of the resulting Gaussian functions in the signal region 
$1.3 < m < 1.8\,\gevcc$ yields an estimate of the $\eetoqq$ background contributions.

\begin{figure*}[hbt]
\begin{center}
\begin{minipage}{.49\textwidth}
\epsfig{file=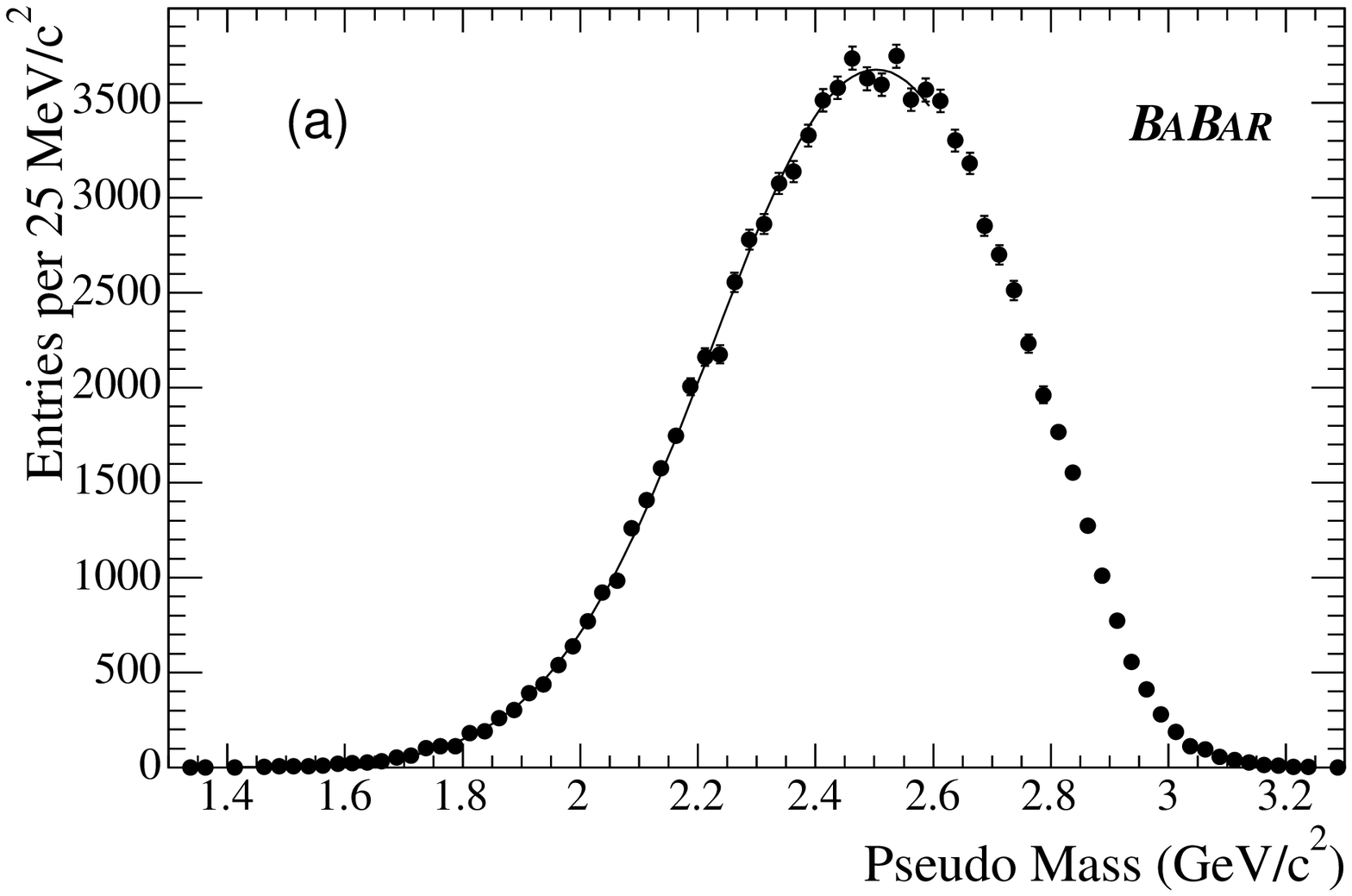,width=1.0\textwidth,height=5cm}
\end{minipage}
\begin{minipage}{.49\textwidth}
\epsfig{file=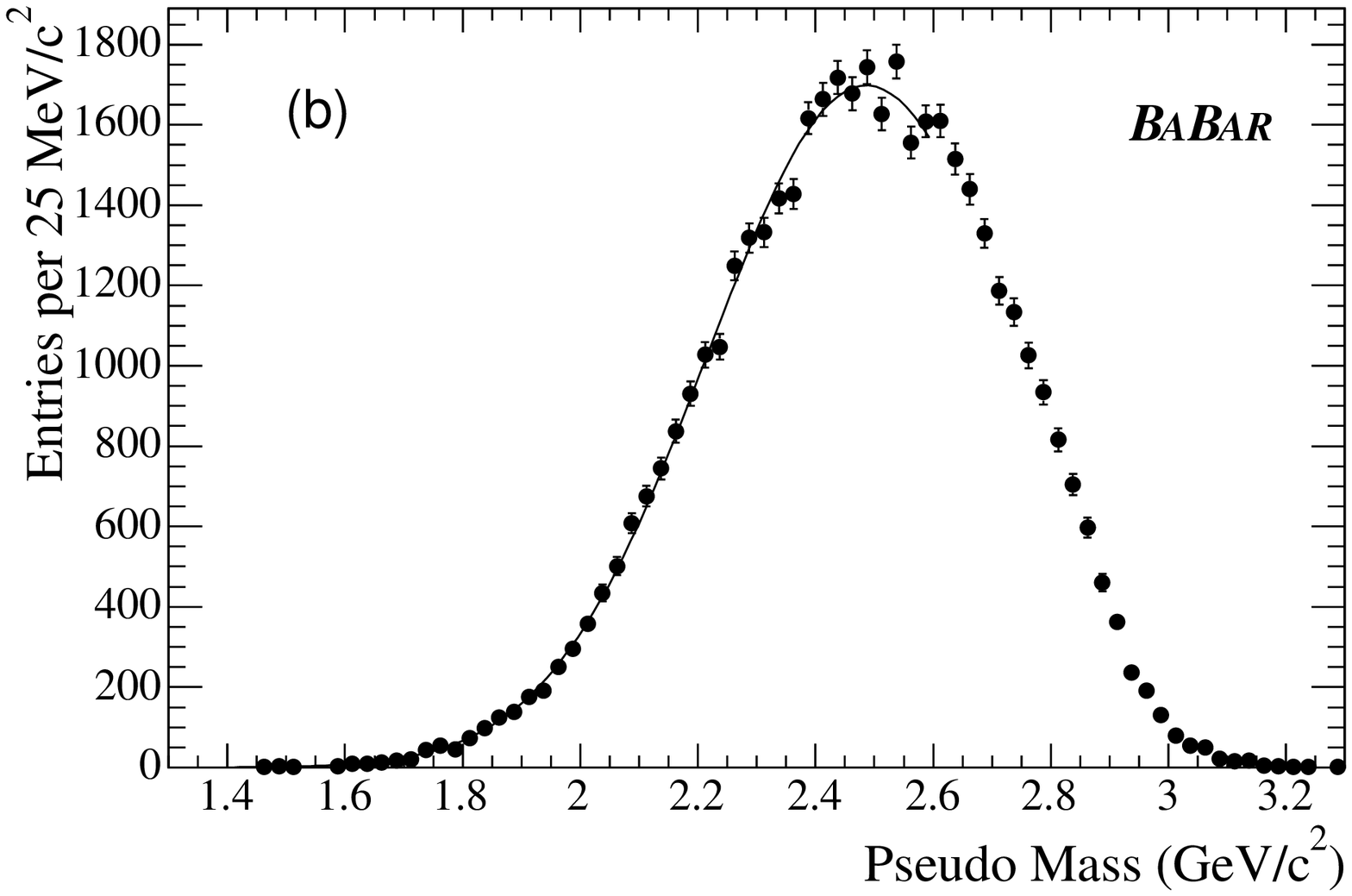,width=1.0\textwidth,height=5cm}
\end{minipage}
\begin{minipage}{.49\textwidth}
\epsfig{file=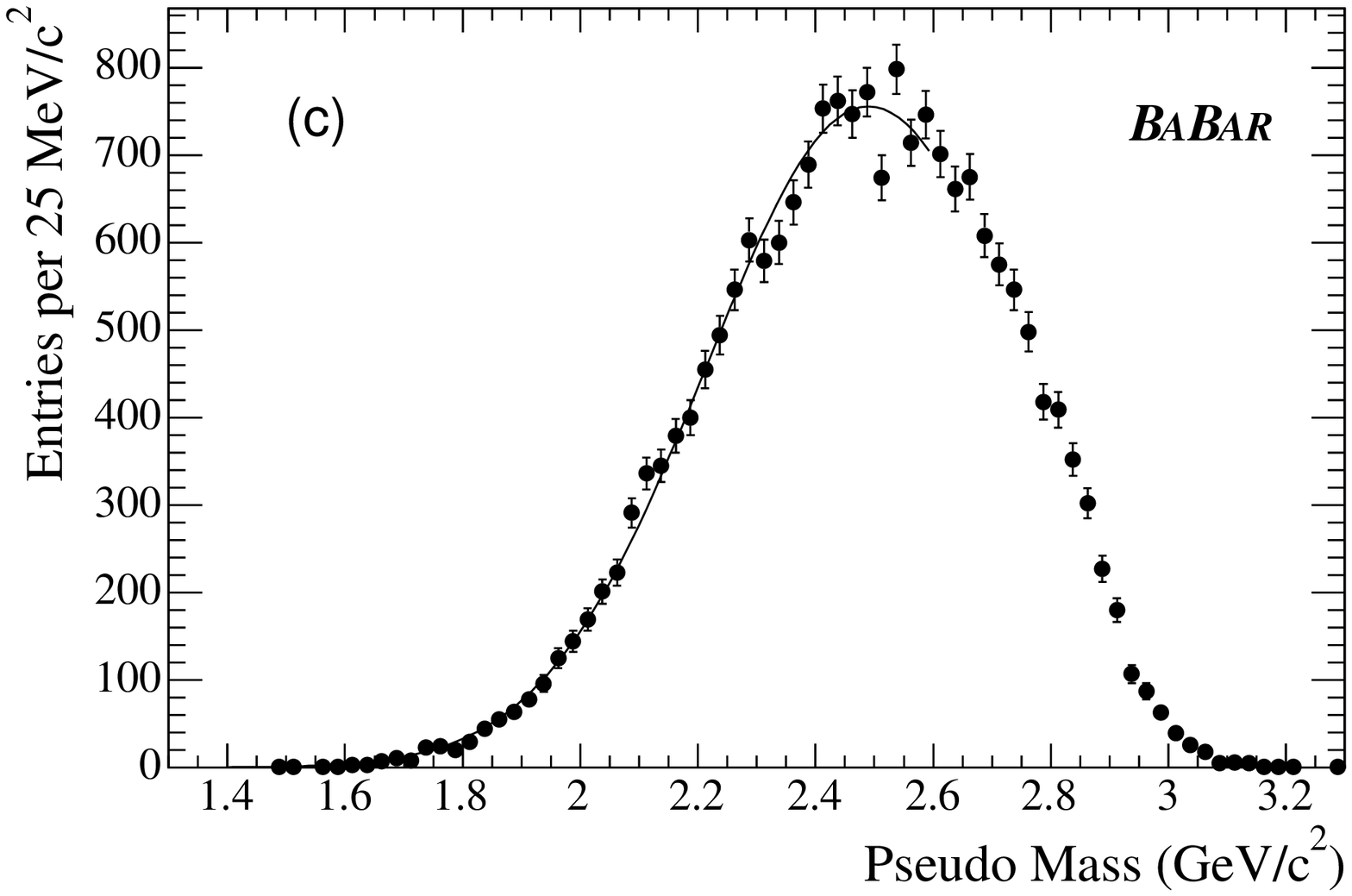,width=1.0\textwidth,height=5cm}
\end{minipage}
\begin{minipage}{.49\textwidth}
\epsfig{file=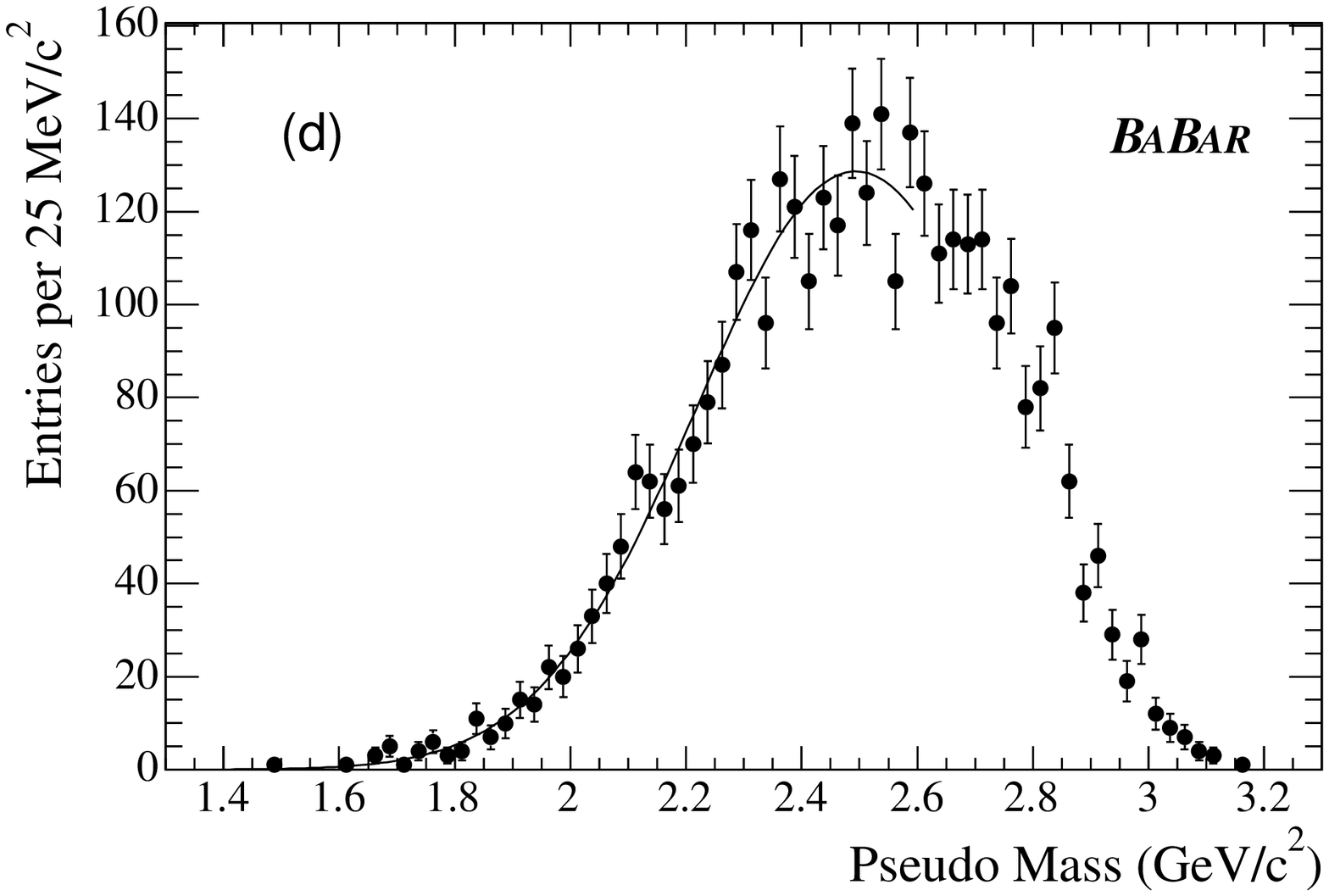,width=1.0\textwidth,height=5cm}
\end{minipage}
\end{center}
\caption{Pseudo mass distributions for 1-7 topology data. (a) after
1-7 topology selection; (b) after particle identification; (c) after conversion veto;
(d) after 1-prong tag.\label{fig:massdata}}
\end{figure*}

Table~\ref{tab:17} lists the number of expected $\tau$ and $\eetoqq$ background events
in the signal region after the four selection steps. In Figure~\ref{fig:massdata} 
(a) -- (d) we show the tau-background-subtracted pseudo mass data
distribution after the four selection steps. 
Although entries below $1.8\,\gevcc$ are shown here, these events were
hidden during the development of our analysis to avoid experimenter bias.
Overlaid in the individual figures are the fit results to the
pseudo mass spectra in the range of 1.8 to $2.6\,\gevcc$.

\begin{figure*}[tb]
\begin{center}
\begin{minipage}{.31\textwidth}
\epsfig{file=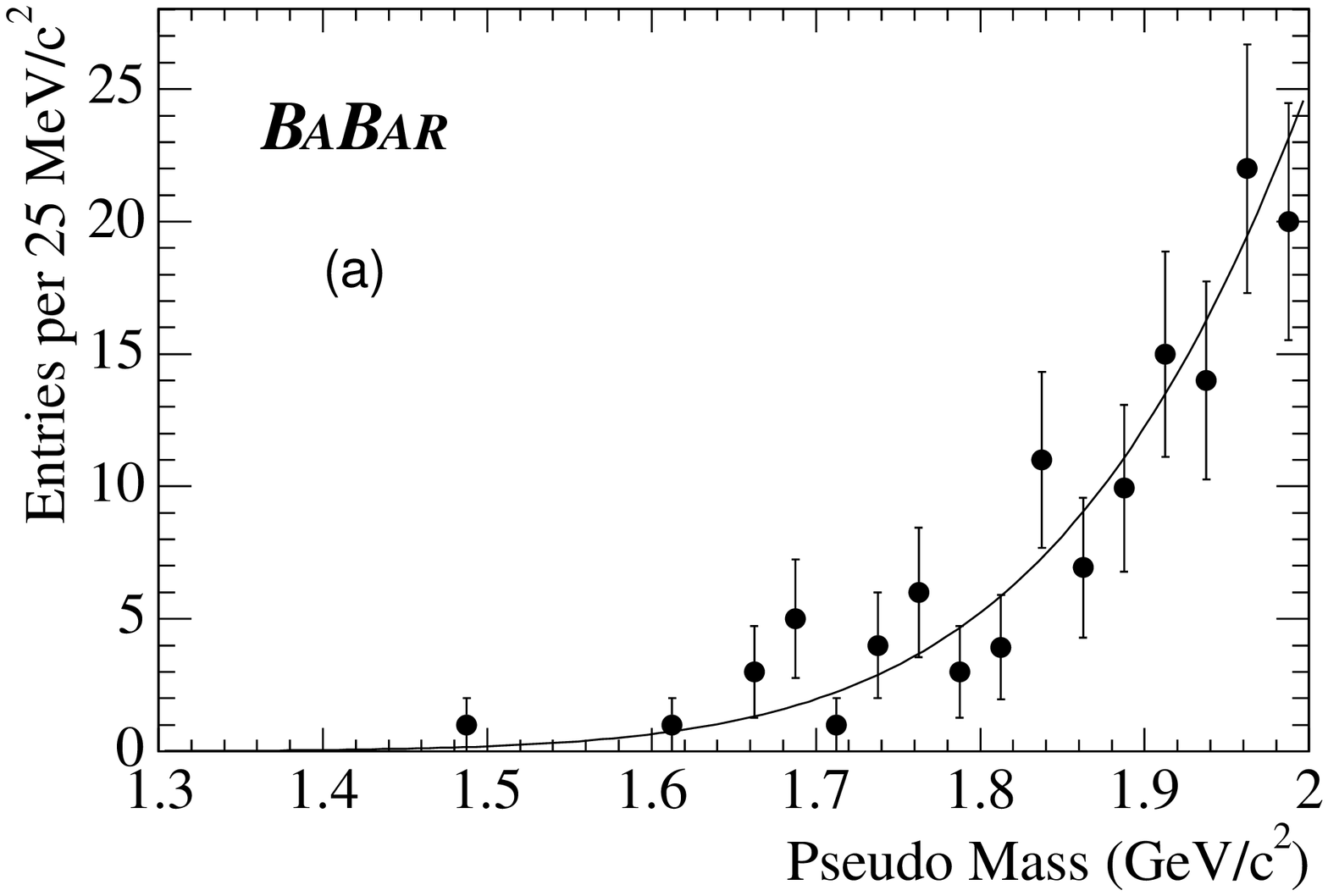,width=1.0\textwidth}
\end{minipage}
\begin{minipage}{.31\textwidth}
\epsfig{file=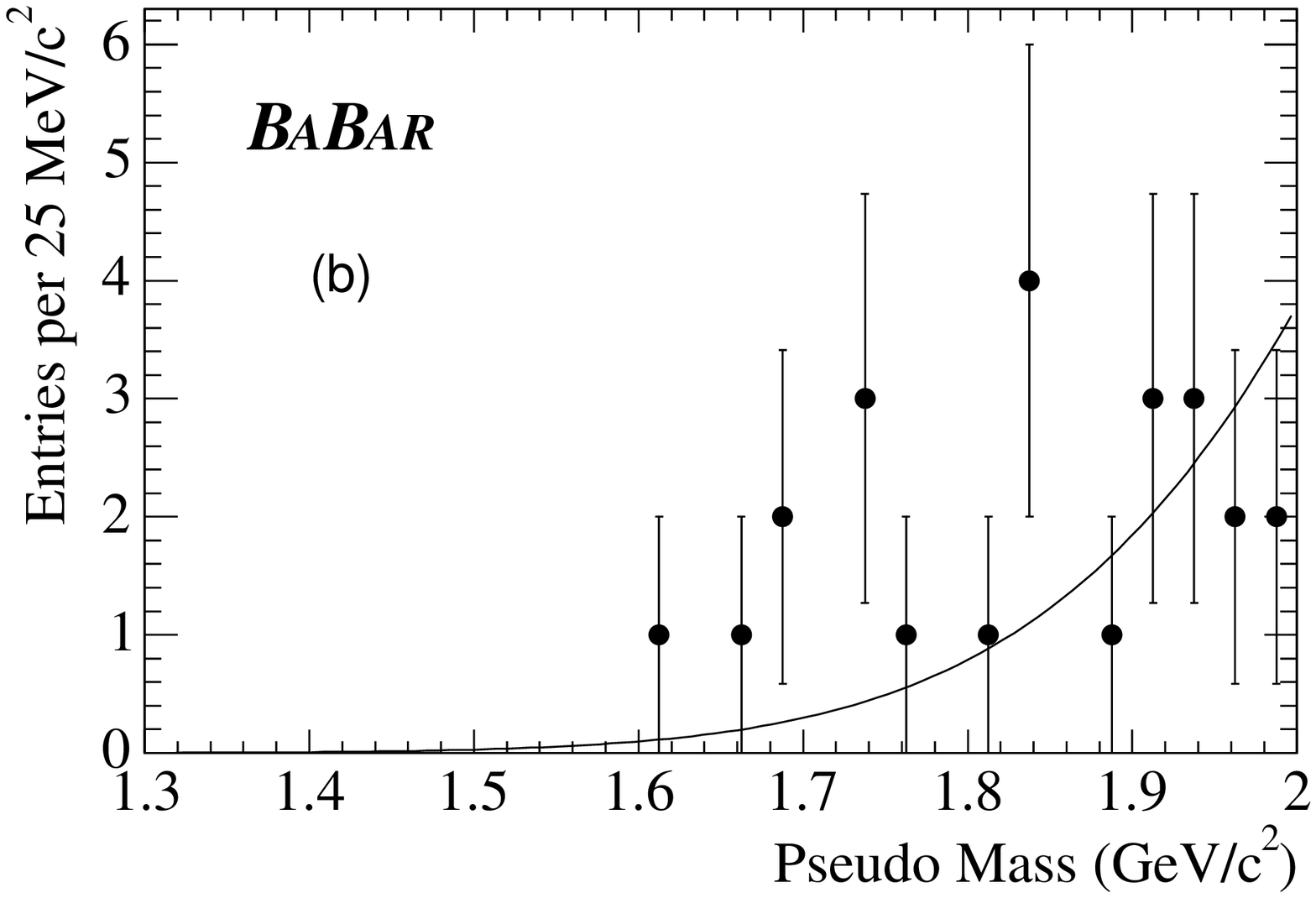,width=1.0\textwidth}
\end{minipage}
\begin{minipage}{.31\textwidth}
\epsfig{file=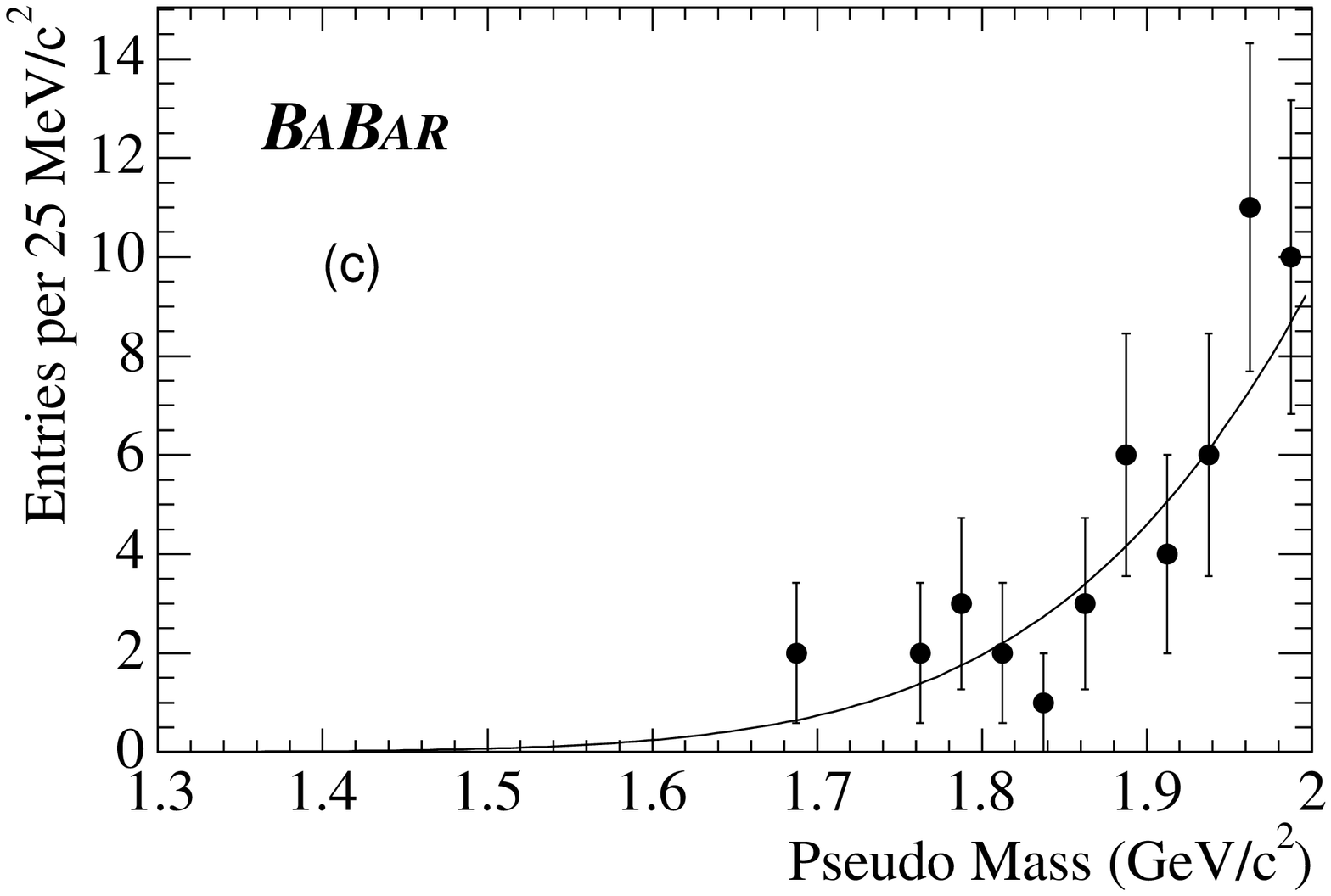,width=1.0\textwidth}
\end{minipage}
\caption{Details of the data pseudo mass spectra.
(a) $\tausevenall$,
(b) $\tauseven$,
(c) $\tausevenpinull$ event
candidates.\label{fig:data_detail}}
\end{center}
\end{figure*}

After the final selection we observe 24 events in the data
(see Figure~\ref{fig:data_detail} (a)\,)
with a total number of predicted background from $\tau$ and $\eetoqq$ events of
$21.6\pm1.2$. The statistical error of the $\eetoqq$ background estimate is derived 
from the statistical uncertainties of the parameters in the Gaussian fit.

\begin{table}[tb]
\begin{center}
\caption{Predicted and observed number of events in the signal region
of $1.3<m<1.8\gevcc$. 
The $\tau$ background yield is obtained from the simulation, 
and its error reflects finite Monte Carlo statistics. The $\eetoqq$ yield 
is determined by fitting the data in the pseudo mass sideband, and 
its error results from data statistics.\label{tab:17}\\}
\begin{tabular}{l|c|c|c}
\hline
\hline
 & $\tau$ Bg.&$\eetoqq$ Bg. & Observed\\
\hline
1-7 Topology	&$128\pm13$ & $574\pm21$   & 695  \\
Particle ID	&$28\pm6$   & $241\pm10$   & 244  \\
Conv.\,Veto     &$2.4\pm1.3$& $119\pm5$    & 104  \\
1-Prong\,Tag 	&$1.3\pm1.0$& $20.3\pm0.7$ &  24  \\
\hline
\hline
\end{tabular}
\end{center}
\end{table}

To validate the $\eetoqq$ background estimation
we use 1-8 topology data~\footnote{The event selection of the 1-7
  topology was modified accordingly, to accommodate eight tracks on the
  signal side.}, which has negligible expected signal contributions.
We fit the pseudo mass distribution between 1.8 and $2.6\,\gevcc$,
integrate the fit function in the pseudo mass region 1.3 to $1.8\,\gevcc$, and
compare this with the number of events found in data (see Table~\ref{tab:18},
columns two and four).
In addition, we repeat the fit with mean and sigma floating (``free''
fit) in each
individual distribution to compare the expected numbers of events
(Table~\ref{tab:18}, column three).
As a final cross check, fits with the Gaussian mean and width floating were 
performed for the 1-7 topology data,
and the results are included in Table~\ref{tab:18}, column five.
We note the very good agreement of the fits with fixed or floating
Gaussian parameters.

The systematic error on the number of expected background events 
is $\pm0.4$ events, derived from
variations of the fit range and shape of the extrapolation function.
The total number of expected background events is $21.6\pm1.3$.\\
\begin{table}[tb]
\begin{center}
\caption{Observed and predicted number of $\eetoqq$ background events in
  the 1-8 and 1-7 topology data for the different selection
  criteria.\label{tab:18}}
\begin{tabular}{l|c|c|c||c}
\hline
\hline
 &1-8 (fixed) &1-8 (free) & 1-8 obs.&1-7 (free)\\
\hline
Topology	&$19.0\pm2.7$& $19.0\pm2.7$ & 23   &$574\pm21$ \\
Particle Id	&$12.2\pm1.6$& $11.2\pm2.0$ & 10   &$222\pm19$\\
Conv.\,Veto     &$2.7\pm0.3$& $2.6\pm1.4$  &  1   &$126\pm18$\\
1-Pr.\,Tag 	&$0.5\pm0.1$& $0.3\pm0.5$  &  0   &$20.2\pm7.7$\\
\hline
\hline
\end{tabular}
\end{center}
\end{table}

We calculate the branching fraction of the 
$\tau^{-} \rightarrow 4 \pi^{-} 3 \pi^{+} (\pi^{0}) \nu_{\tau}$
decay based on the following likelihood function, which 
convolves a Poisson
distribution with two Gaussian resolution functions, accounting 
for the uncertainties in the background and in the efficiency:
\begin{equation}
{\cal L}(\BR; n, \hat b, \hat f, b, f) = 
\frac{\mu^n e^{-\mu}} {n!} 
\frac{1} {2\pi\sigma_b\sigma_f}
e^{
-\frac{1}{2}
\left( \frac{\hat b - b}{\sigma_b} \right)^2 -
\frac{1}{2}
\left( \frac{\hat f - f}{\sigma_f} \right)^2
},
\label{func:like}\end{equation}
where \BR\ denotes the branching fraction of 
$\tau^{-} \rightarrow 4 \pi^{-} 3 \pi^{+} (\pi^{0}) \nu_{\tau}$, 
$f=2N_{\tau\tau}\epsilon$, 
$\epsilon$ is the signal efficiency,
$\mu=\langle n\rangle=\hat f\BR+\hat b$, with
$n$ the number of observed events,
and $b$ and $\sigma_b$ are the expected background and its error.
$\sigma_f$ incorporates the errors on the signal efficiency and the 
number of $\tau$ pair events.\\

The likelihood function is maximized with respect to the branching fraction $\BR$,
$\hat f$ and $\hat b$, and the following numerical value for the branching fraction
 is obtained by MINUIT\cite{James:1975dr}:
\begin{equation}
\BR(\tau^{-} \rightarrow 4 \pi^{-} 3 \pi^{+} (\pi^{0}) \nu_{\tau}) 
= (0.7^{+1.4}_{-1.3})\times10^{-7}.
\nonumber\end{equation}

Since we have no evidence for a signal we compute a Bayesian upper limit using a uniform prior in the
branching fraction, the background, and the efficiency. This is done by integrating
out $\hat f$ and $\hat b$ in the likelihood function and plotting ${\cal L}$ as function of $\BR$.
In this way we normalize the distribution to unity and get as the result of this analysis an upper limit 
at the point where the integral reaches $0.9$:
\begin{equation}
\BR(\tau^{-} \rightarrow 4 \pi^{-} 3 \pi^{+} (\pi^{0}) \nu_{\tau}) 
< 3.0\times10^{-7}\ \rm(at\ 90\,\%\ CL).
\nonumber\end{equation}

With the same approach, setting the number of observed events $N_{obs}$ to 
the expected number of background events of $N_{exp} = 21.6$,
we calculate the sensitivity of the analysis to be
$\BR^{N_{obs}\equiv N_{exp}}(\tau^{-} \rightarrow 4 \pi^{-} 3 \pi^{+} (\pi^{0}) \nu_{\tau}) <
2.5 \times 10^{-7}$ at 90\,\% CL.\\

In addition to this inclusive result, we set limits on the branching fractions 
of the exclusive decay modes $\tauseven$ and $\tausevenpinull$. 
To select $\tauseven$ candidates in the inclusive sample, we require the number of 
photons on the signal side to be zero. This yields a signal efficiency for
$\tauseven$ decays of $(5.5\pm0.3)\,\%$ while reducing the expected
generic $\tau$ decay and $\eetoqq$ background to $3.9\pm0.8$ events.
The decays $\tausevenpinull$ are treated as background in this case
and have a reconstruction efficiency of $(0.8\pm0.1)\,\%$.
Reversing this selection by demanding at least one reconstructed $\pi^0$
on the signal side yields a signal efficiency for the
$\tausevenpinull$ decay of $(3.6\pm0.3)\,\%$ and an expected generic
$\tau$ decay and $\eetoqq$ background of $8.2\pm0.5$ events. 
In this case, the reconstruction efficiency for $\tauseven$ is $(0.3\pm0.0)\,\%$.
The systematic uncertainties of the $\tauseven$ mode are identical to
the inclusive measurement already discussed above. For the $\tausevenpinull$
mode, an additional uncertainty of $5.0\,\%$ on the efficiency of the $\pi^0$ reconstruction 
is taken into account.\\

The likelihood function~(\ref{func:like}) was modified
to accommodate one exclusive mode acting as background
for the other.
We observe eight events in the $\tauseven$ signal region
and seven events in $\tausevenpinull$.
We calculate the following numerical values for the branching fractions:
\begin{eqnarray}
\BR(\tauseven) & = & (2.0^{+1.5}_{-1.2})\times10^{-7} \nonumber\\
\BR(\tausevenpinull) & = & (-1.0\pm1.8)\times10^{-7}.
\nonumber
\end{eqnarray}

Without evidence for a signal we compute Bayesian upper limits using uniform priors in the
branching fractions, the backgrounds, and the efficiencies, of 
\begin{eqnarray}
\BR(\tauseven) & < & 4.3\times10^{-7}\ \rm(at\ 90\,\%\ CL) \nonumber\\
\BR(\tausevenpinull) & < & 2.5\times10^{-7}\ \rm(at\ 90\,\%\ CL).
\nonumber\end{eqnarray}

With the same approach, setting the number of observed events $N_{obs}$ to
the expected number of background events, we calculate the sensitivities
$\BR^{N_{obs}\equiv N_{exp}}(\tauseven) < 2.2 \times 10^{-7}$
and
$\BR^{N_{obs}\equiv N_{exp}}(\tausevenpinull) < 4.2 \times 10^{-7}$.
Figures~\ref{fig:data_detail} (b) and (c) show details of the data pseudo mass
spectra with an overlay of the expected background distributions.

This analysis improves the existing experimental limits by an order of
magnitude for the inclusive mode, but is still several orders of
magnitude larger than the theoretical prediction.
The exclusive decays are reported for the first time and are
consistent with the inclusive result.\\

We are grateful for the 
extraordinary contributions of our \pep2\ colleagues in
achieving the excellent luminosity and machine conditions
that have made this work possible.
The success of this project also relies critically on the 
expertise and dedication of the computing organizations that 
support \babar.
The collaborating institutions wish to thank 
SLAC for its support and the kind hospitality extended to them. 
This work is supported by the
US Department of Energy
and National Science Foundation, the
Natural Sciences and Engineering Research Council (Canada),
Institute of High Energy Physics (China), the
Commissariat \`a l'Energie Atomique and
Institut National de Physique Nucl\'eaire et de Physique des Particules
(France), the
Bundesministerium f\"ur Bildung und Forschung and
Deutsche Forschungsgemeinschaft
(Germany), the
Istituto Nazionale di Fisica Nucleare (Italy),
the Foundation for Fundamental Research on Matter (The Netherlands),
the Research Council of Norway, the
Ministry of Science and Technology of the Russian Federation, and the
Particle Physics and Astronomy Research Council (United Kingdom). 
Individuals have received support from 
CONACyT (Mexico),
the A. P. Sloan Foundation, 
the Research Corporation,
and the Alexander von Humboldt Foundation.

\bibliography{ref}  

\end{document}